\documentclass[12pt]{article}
\usepackage{amsmath,amssymb}
\begin{document}
\begin{flushright} SU-4252-815
\\
\end{flushright}
\begin{center}
\vskip 3em
{\LARGE UV-IR Mixing in Non-Commutative Plane}
\vskip 2em
{\large A. P. Balachandran$^\dagger$, A. Pinzul$^\dagger$ and B. A. Qureshi$^\dagger$
\footnote[1]{E-mails: bal@phy.syr.edu, apinzul@phy.syr.edu, bqureshi@phy.syr.edu}
\\[2em]}
\em{\oddsidemargin 0 mm
$^\dagger$Department of Physics, Syracuse University,
Syracuse, NY 13244-1130, USA}\\
\end{center}
\vskip 1em
\begin{abstract}
Poincar\'e-invariant quantum field theories can be formulated on
non-commutative planes if the coproduct on the Poincar\'e group is
suitably deformed \cite{Dimitrijevic:2004rf, Chaichian:2004za}.(See also especially Oeckl \cite{Oeckl:1999jun},\cite{Oeckl:2000mar} and Grosse et al.\cite{Grosse:2001mar}) As
shown in \cite{Balachandran:2005eb}, this important result of
these authors implies modification of free field commutation and
anti-commutation relations and striking phenomenological
consequences such as violations of Pauli principle
\cite{Balachandran:2005eb,Bal3}. In this paper we prove that with
these modifications, UV-IR mixing disappears to all orders in
perturbation theory from the $S$-Matrix. This result is in agreement with the previous results of Oeckl \cite{Oeckl:2000mar}.
\end{abstract}
\newpage

\setcounter{footnote}{0}

\section{Introduction}

The non-commutative Groenwold-Moyal plane is the algebra
$\mathcal{A}_\theta (\mathbb{R}^{d+1})$ of functions on
$\mathbb{R}^{d+1}$ with the $*$-product as the multiplication law.
The latter is defined as follows.

If $\alpha,\beta\in\mathcal{A}_\theta(\mathbb{R}^{d+1})$, then
\begin{equation}
\alpha\ast_{\theta}\beta\ (x)=(\alpha\
e^{\frac{i}{2}{\overleftarrow
\partial}_{\mu}\theta^{\mu\nu}{\overrightarrow \partial}_{\nu}}\
\beta)(x)\ ,
\end{equation}
\begin{equation}
\theta^{\mu\nu}=-\,\theta^{\nu\mu}\in\mathbb{R}\ ,
\end{equation}
\begin{equation}
x=(x^0,x^1,\dots,x^d). \nonumber
\end{equation}
Here $x^0$ is the time coordinate, and the rest are spatial coordinates.

Henceforth, we will write $\alpha\ast_{\theta}\beta$ as $\alpha\ast\beta$.

The appearance of constants $\theta^{\mu\nu}$ would at first sight
suggest that the diffeomorphism group
$\it{Diff}\,(\mathbb{R}^{d+1})$ of $\mathbb{R}^{d+1}$, and in
particular its Poincar\'e subgroup is not an automorphism of
$\mathcal{A}_\theta(\mathbb{R}^{d+1})$. But the work of
\cite{Dimitrijevic:2004rf} and \cite{Chaichian:2004za} (and the earlier work of \cite{Oeckl:2000mar} and \cite{Grosse:2001mar}) have shown
that this appearance is false. Thus there exists a deformed coproduct
on $\it{Diff}\,(\mathbb{R}^{d+1})$ which depends on $\theta^{\mu\nu}$. With this deformation,
$\it{Diff}\,(\mathbb{R}^{d+1})$ does act as the automorphism group
of $\mathcal{A}_{\theta}(\mathbb{R}^{d+1})$.

In \cite{Balachandran:2005eb} (and the earlier work of \cite{Oeckl:2000mar} and \cite{Grosse:2001mar}), it was shown that the standard
commutation relations are not compatible with the deformed action
of Poincar\'e group. Rather they too have to be deformed. If
$a(p)$ is the annihilation operator of a free field for momentum
$p$, then for example,
\begin{equation}
a(p)\,a(q)=\eta\,e^{ip_\mu\theta^{\mu\nu}q_\nu}\ a(q)\,a(p),
\end{equation}
where $\eta$ is a Lorentz-invariant function of $p$ and $q$. The
choices $\eta=\pm 1$ correspond, for $\theta=0$, to bosons and
fermions.

There are similar relations involving $a(p)^\dagger$'s as well.
All of them follow from the relations
\begin{equation}
a(p)=c(p)e^{+\frac{i}{2}p_\mu\theta^{\mu\nu}P_\nu},
\label{eq:map}
\end{equation}
\begin{equation}
a(p)^\dagger=e^{-\frac{i}{2}p_\mu\theta^{\mu\nu}P_\nu}\,c(p)^\dagger
\end{equation}
where $c(p)$ and $c(p)^\dagger$ are the standard oscillators
$a(p)\mid_{\theta=0}\ ,\ a(p)^\dagger\mid_{\theta=0}$ for
$\theta=0$, and $P_\mu$ is the translation generator:
\begin{equation}
P_\mu=\int d\mu(p)\,p_\mu\,c(p)^\dagger\,c(p)\ =\ \int
d\mu(p)p_\mu\,a(p)^\dagger\,a(p).
\end{equation}
$d\mu(p)$ here is the Poincar\'e-invariant measure. For a spin 0
field of mass m,
\begin{equation}
d\mu(p)\ =\ \frac{d^3p}{2p_0}\ ,\ p_0=\left|\sqrt{{\overrightarrow
p}^2+m^2}\right|
\end{equation}
There are striking consequences of the deformed commutation
relation \cite{Balachandran:2005eb} such as the existence of
Pauli-forbidden levels and attendant phenomenology \cite{Bal3}. In
this note, we show another striking result: Non-planar graphs and
UV-IR mixing completely disappear from the S-matrix $S_\theta$
because of the deformed statistics. $S_\theta$ is in fact
independent of $\theta^{\mu\nu}$ so that  $S_\theta\ =\ S_0$. This
does not mean that scattering amplitudes are independent of
$\theta$, as the in- and out- state vectors are different, being
subject to deformed statistics.

Our treatment here covers both time-space and space-space
noncommutativity. In the former case, although there were initial
claims of loss of unitarity, the work of Doplicher et
al.\cite{Doplicher:1994zv} showed how to construct unitary
theories. These ideas were subsequently applied to construct
unitary quantum mechanics as well
\cite{Balachandran:2004rq,Balachandran:2004yh}. So there is no
good theoretical reason to set $\theta^{0i}=0$. The work we
present here is quite general as regards the choice of
$\theta^{\mu\nu}$, allowing also the choice $\theta^{0i}\ne 0$.

We present the calculations for a real scalar field with the
interaction
$$
\phi_\ast^n\,:=\,\phi\,\ast\,\phi\,\ast\,\dots\,\phi\ (n\ge 2)\ .
$$
The generality of the results will be evident from this example.

There is considerable overlap of the results of this work with those of Oeckl \cite{Oeckl:2000mar}. He too uses nontrivial twisted statistics, but does not use Poincar\'e symmetry implemented with a twisted coproduct\cite{Dimitrijevic:2004rf, Chaichian:2004za}. In contrast, our previous work \cite{Balachandran:2005eb} deduced  twisted statistics from Poincar\'e invariance. Oeckl then deduces an expression for the $n$-point function in agreement with ours. His derivation is based on braided quantum field theory developed by him \cite{Oeckl:1999jun}. Its relation to our approach awaits clarification. But we point out that once the appropriately twisted spacetime algebra and statistics are accepted as axioms, both Oeckl and us get the same final answer without ever invoking Poincar\'e invariance or any other spacetime symmetry except translations.

\section{The Model}

The free scalar field $\phi$ of mass $m$ in the Moyal plane has
the Fourier expansion
\begin{equation}
\phi(x)=\ \int d\mu(p)\,[a(p) e^{ip\cdot x}+a(p)^\dagger e^{-ip\cdot x}]\ ,
\end{equation}
\[p_0=\sqrt{{\overrightarrow p}^2+m^2}.\]
The interaction Hamiltonian, in the interaction representation, is taken to be
\begin{equation}
H_I(x_0)=\ \lambda \int d^d x\ :\phi_\ast ^n :
\end{equation}
where $:\ \ \ :$ denote normal ordering of $a(p)$'s and
${a(p)^\dagger}$'s.

The operator $H_I (x_0)$ is self-adjoint for any choice of
$\theta^{\mu\nu}$, even with time-space noncommutativity. Hence
the S-matrix
\begin{eqnarray}
\lefteqn{ S_\theta = T \, \exp\left(\ -i\ \int d x_0 \, H_I(x_0)\right) {}} \nonumber \\
& & {} = T \, \exp\left(\ -i\ \int d^{d+1} x \ :\ \phi_\ast^n (x)
\ :\right) \nonumber
\end{eqnarray}
is unitary. We will now show that $S_\theta$ is independent of
$\theta$. That means in particular that there is no UV-IR mixing.

Let $e_p$ be the plane wave of momentum $p$:  $e_p=e^{ip.x}$. The
$\ast$-product of plane waves is simple :
\begin{equation}
e_p\ \ast \ e_q= \ e^{-\frac{i}{2}p_\mu\theta^{\mu\nu}q_\nu}\
e_{p+q}\ .
\end{equation}
Let us introduce the notation \[a(p)^\dagger=a(-p)\] where $p_0$ is also reversed by the dagger. Then
\begin{equation}
\phi = \ \int d\mu(p) \ [a(p)e_p\ +\ a(-p)e_{-p}]\ .
\end{equation}

\section{The Proof}

\vspace{.5cm}

{\it i)} ${n=2}$

\vspace{.5cm}

First consider $n=2$, just as an example. Then the
$\it{O}(\lambda)$ term of $ S_\theta$ is
\begin{equation}
S_\theta ^{(1)}=\ -i\lambda \ \int d^{d+1}x\ \ :\,\phi\,\ast \,
\phi\,:(x)\ .
\end{equation}
A typical term in $\phi\ast\phi$ is\footnote{Here we have used
$e_p\ \ast \ e_q= \ e^{\frac{i}{2}p_\mu\theta^{\mu\nu}q_\nu}\
e_{p+q}$, which requires replacing $\theta^{\mu\nu}$ by
$-\theta^{\mu\nu}$ in (1). The reason for this change is explained
in \cite{Balachandran:2005eb} after Eq.(2.33).}
\begin{equation}
a(p)a(q)\ e_p\ast e_q =\ a(p)a(q)\
e^{\frac{i}{2}p_\mu\theta^{\mu\nu}q_\nu}\ e_{p+q}\ . \label{eq:A}
\end{equation}
Substituting from (\ref{eq:map}), we get
\begin{eqnarray}
\textrm{R.H.S of (\ref{eq:A})}& = &
c(p)e^{\frac{i}{2}p_\mu\theta^{\mu\nu}P_\nu}\
c(q)e^{\frac{i}{2}q_\nu\theta^{\mu\nu}P_\nu}\
e^{\frac{i}{2}p_\mu\theta^{\mu\nu}q_\nu}\ e_{p+q}\nonumber\\
& = &c(p)\ c(q) e^{-\frac{i}{2} p_\mu\theta^{\mu\nu}q_\nu}\
e^{\frac{i}{2} p_\mu\theta^{\mu\nu}q_\nu}\
e^{\frac{i}{2}(p+q)_\mu\theta^{\mu\nu}P_\nu}\ e_{p+q} \nonumber\\
& &\left(\mbox{since }[P_\nu,c(q)]=-q_\nu\,c(q)\right)\nonumber\\
& = &\ c(p)c(q)\ e_{p+q}
e^{\frac{i}{2}(p+q)_\mu\theta^{\mu\nu}P_\nu}\ .
\end{eqnarray}
Note how the phases $e^{\mp\frac{i}{2}p_\mu\theta^{\mu\nu}q_\nu}$
cancel.

Using \[\partial_\mu\,e_{p+q}\ =\ i(p+q)_\mu\ e_{p+q},\] we can
write this as \[c(p)c(q)\,e_{p+q}\ e^{\frac{1}{2}{\overleftarrow
\partial}_\mu\theta^{\mu\nu}P_\nu}.\] Hence
\begin{eqnarray}
\lefteqn{-i\lambda\ \int d^{d+1}x\ \ :\phi\ \ast\
\phi:(x){}}\nonumber \\ & & {} =\ -i\lambda \ \int d^{d+1}x\ \
:\phi^2:(x)\ \ e^{\frac{1}{2}{\overleftarrow
\partial}_\mu\theta^{\mu\nu}P_\nu } .
\end{eqnarray}
Expanding the exponential, integrating and discarding the surface
terms, we find that
$$
-i\lambda\ \int d^{d+1}x\ :\phi\ast\phi:(x)\ =\ -i\lambda\ \int
d^{d+1}x\ :\phi^2:(x)
$$
is independent of $\theta^{\mu\nu}$.

The only delicate issue here concerns the surface term. Here and
in what follows, we will assume that such surface terms vanish. In
the absence of long range forces, the assumption should be
correct.

Next consider the $\it{O}(\lambda^2)$ term
\begin{eqnarray}
S_\theta^{(2)}&=& \frac{(-i\lambda)^2}{2!}\int
d^{d+1}x_1\,d^{d+1}x_2\ \left\{\theta(x_{10}-x_{20})
{}:\phi\ast\phi:(x_1)\,:\phi\ast\phi:(x_2)\right.\nonumber \\
& & +\ \ \left.(x_1\leftrightarrow x_2)\right\}\ .
\end{eqnarray}
A typical term in
$\theta(x_{10}-x_{20}):\phi\ast\phi:(x_1)\,:\phi\ast\phi:(x_2)$ is
\begin{equation}
\theta(x_{10}-x_{20}):a(p_1)a(q_1):e_{p_1}\ast
e_{q_1}\,(x_1):a(p_2)a(q_2):e_{p_2}\ast e_{q_2}\,(x_2) \nonumber
\end{equation}
\begin{eqnarray}
\lefteqn{=\theta(x_{10}-x_{20}):c(p_1)c(q_1):
e_{p_1+q_1}(x_1)e^{+\frac{i}{2}(p_1+q_1)_\mu\theta^{\mu\nu}P_\nu}} \nonumber \\
& &
:c(p_2)c(q_2):e_{p_2+q_2}(x_2)e^{+\frac{i}{2}(p_2+q_2)_\mu\theta^{\mu\nu}P_\nu}\nonumber
\end{eqnarray}
\begin{eqnarray}
\lefteqn{=\theta(x_{10}-x_{20})\left\{:c(p_1)c(p_2)::c(q_1)c(q_2):
e^{-\frac{i}{2}(p_1+q_1)_\mu\theta^{\mu\nu}(p_2+q_2)_\nu}\right.}\nonumber \\
& & \left.\left[e_{p_1+q_1}(x_1)\ e_{p_2+q_2}(x_2)\
e^{+\frac{1}{2}(\frac{{\overleftarrow \partial}}{\partial
x_{1\mu}}+\frac{{\overleftarrow \partial}}{\partial
x_{2\mu}})\theta^{\mu\nu}P_\nu}\ \right]\right\} \label{eq:big}
\end{eqnarray}
where the differentials act only on $e_{p_1+q_1},\ e_{p_2+q_2}$
and phases involving just $p_\mu$ and $q_\mu$ cancelling out as
before.

Note first that by energy-momentum conservation [enforced by
integration over $x_1+x_2$ and the resultant $\delta^{d+1}(\sum
p_i)$], we can set $p_2+q_2=-p_1-q_1$. Hence we can set
\[ e^{-\frac{i}{2}(p_1+q_1)_\mu\theta^{\mu\nu}(p_2+q_2)_\nu}\,=\,1.\]
Next note that since
$$
(\frac{ \partial}{\partial x_{10}}+\frac{
\partial}{\partial x_{20}})\theta(x_{10}-x_{20})=0,
$$
we can in fact allow $\frac{{\overleftarrow \partial}}{\partial
x_{10}}+\frac{{\overleftarrow \partial}}{\partial x_{20}}$ to act
on the $\theta$-function as well. But then all terms involving
$\theta^{\mu\nu}$ in the power series expansion of the exponential
are total differentials and vanish upon integrating over
$d^{d+1}x_1\,d^{d+1}x_2$. Thus
\[S_\theta^{(2)}=S_0^{(2)}.\]
Similar calculations show that $S_\theta$ is independent of
$\theta^{\mu\nu}$ exactly, to all orders in $\theta^{\mu\nu}$.
\[S_\theta\ =\ S_0\qquad \textrm{for}\quad n=2.\]

\vspace{.5cm}

{\it ii)} Generic $n$

\vspace{.5cm}

The typical term in
$$
:\underbrace{\phi\ast\phi\ast\dots\ast\phi}_{n-terms}:(x)
$$ is
$$
:a(p)a(q)\dots a(s):\ e_p\ast e_q\ast \dots \ast e_s\ (x)
$$ which too
simplifies to
$$
:c(p)c(q)\dots c(s):\ e_{p+q+\dots +s}\,(x)\
e^{+\frac{i}{2}(p+q+\dots +s)_\mu\theta^{\mu\nu}P_\nu}$$ for any
$n$. Hence, we find to $\it{O}(\lambda)$, for any $n$, as before
that
$$
S_\theta^{(1)}\ =\ S_0\ .
$$

The proof to higher orders is similar. Thus to
$\it{O}(\lambda^2)$, (\ref{eq:big}) is replaced by
\begin{eqnarray}
& &=\theta(x_{10}-x_{20})\Big\{\ :c(p_1^{(1)})\cdots
c(p_1^{(n)}):\ \ :c(p_2^{(1)})\cdots c(p_2^{(n)}):\
e^{-\frac{i}{2}\Big(\sum_{j=1}^{n}
(p_1^{(j)})_\mu\Big)\theta^{\mu\nu}
\Big(\sum_{k=1}^{n} (p_2^{(k)})_\nu\Big) }{} \nonumber \\
& &\Big[e_{\sum_{j} p_1^{(j)}}\,(x_1)\ \,e_{\sum_{k}
p_2^{(k)}}\,(x_2)\ \,e^{+\frac{1}{2}(\frac{{\overleftarrow
\partial}}{\partial x_{1\mu}}+\frac{{\overleftarrow
\partial}}{\partial x_{2\mu}})\theta^{\mu\nu}P_\nu}\ \Big]\Big\}
\end{eqnarray}
which can again be shown to be independent of $\theta^{\mu\nu}$
using energy-momentum conservation and partial integration.
Therefore
$$
S_\theta^{(2)}\ =\ S_0^{(2)}\ .
$$
This proof extends to all orders so that
$$
S_\theta\ =\ S_0\ .
$$

\vspace{.5cm}

{\bf Acknowledgments}

\vspace{.5cm}

We thank Giampiero Mangano for extensive discussions. The work was
supported by DOE under grant number DE-FG02-85ER40231 and by NSF
under contract number INT9908763.


\begin{thebibliography}{12}




\bibitem{Dimitrijevic:2004rf}
  M.~Dimitrijevic and J.~Wess,
  ``Deformed bialgebra of diffeomorphisms,''
  arXiv:hep-th/0411224;
  P.~Aschieri, C.~Blohmann, M.~Dimitrijevic, F.~Meyer, P.~Schupp and J.~Wess,
  ``A gravity theory on noncommutative spaces,''
  arXiv:hep-th/0504183.

\bibitem{Chaichian:2004za}
  M.~Chaichian, P.~P.~Kulish, K.~Nishijima and A.~Tureanu,
  ``On a Lorentz-invariant interpretation of noncommutative space-time and its
  implications on noncommutative QFT,''
  Phys.\ Lett.\ B {\bf 604}, 98 (2004)
  [arXiv:hep-th/0408069];
  M.~Chaichian, P.~Presnajder and A.~Tureanu,
  ``New concept of relativistic invariance in NC space-time: Twisted  Poincare
  symmetry and its implications,''
  Phys.\ Rev.\ Lett.\  {\bf 94}, 151602 (2005)
  [arXiv:hep-th/0409096].


\bibitem{Oeckl:1999jun}
 Robert Oeckl,
 ``Braided Quantum Field Theory'',
 Commun. Math. Phys. 217 (2001) 451-473
 [arXiv:hep-th/9906225];

\bibitem{Oeckl:2000mar}
 Robert Oeckl,
 ``Untwisting Noncommutative $R^d$ and the Equivalence of Quantum Field
 Theories'',
 Nucl.Phys. B581 (2000) 559-574
 [arXiv:hep-th/0003018].

\bibitem{Grosse:2001mar}
 H. Grosse, J. Madore, H. Steinacker,
 ``Field theory on the $q$-deformed Fuzzy Sphere II: Quantization'',
 J.Geom.Phys. 43(2002) 205-240
 [arXiv:hep-th/0103164].


\bibitem{Balachandran:2005eb}
  A.~P.~Balachandran, G.~Mangano, A.~Pinzul and S.~Vaidya,
  ``Spin and statistics on the Groenwold-Moyal plane: Pauli-forbidden levels
  and transitions,''
  arXiv:hep-th/0508002.

\bibitem{Bal3}
A. P. Balachandran, G. Mangano, A. Pinzul and B. Qureshi (work in progress).

\bibitem{Doplicher:1994zv}
  S.~Doplicher, K.~Fredenhagen and J.~E.~Roberts,
  ``Space-time quantization induced by classical gravity,''
  Phys.\ Lett.\ B {\bf 331}, 39 (1994);
  S.~Doplicher, K.~Fredenhagen and J.~E.~Roberts,
  ``The Quantum structure of space-time at the Planck scale and quantum
  fields,''
  Commun.\ Math.\ Phys.\  {\bf 172}, 187 (1995)
  [arXiv:hep-th/0303037].



\bibitem{Balachandran:2004rq}
  A.~P.~Balachandran, T.~R.~Govindarajan, C.~Molina and P.~Teotonio-Sobrinho,
  ``Unitary quantum physics with time-space noncommutativity,''
  JHEP {\bf 0410}, 072 (2004)
  [arXiv:hep-th/0406125].

\bibitem{Balachandran:2004yh}
  A.~P.~Balachandran, T.~R.~Govindarajan, A.~G.~Martins and P.~Teotonio-Sobrinho,
  ``Time-space noncommutativity: Quantised evolutions,''
  JHEP {\bf 0411}, 068 (2004)
  [arXiv:hep-th/0410067].



\end{thebibliography}
\end{document}